\documentclass[10pt,conference]{IEEEtran}

\usepackage{paralist}
\usepackage{caption}

\usepackage{multirow}
\usepackage{graphicx}
\usepackage{textcomp}
\usepackage{listings}
\usepackage{xspace}
\usepackage[numbers]{natbib}
\usepackage[binary-units,per-mode=symbol,detect-weight=true, detect-family=true]{siunitx}
\newcommand{\tempsydslname}{SB-TemPsy-DSL\xspace}
\newcommand{\app}{ThEodorE\xspace}

\usepackage[inline,shortlabels]{enumitem}
\usepackage{xintexpr}
\usepackage{newtxmath}

\usepackage{url}

\usepackage{breakurl}
\usepackage[breaklinks]{hyperref}

\usepackage{xcolor}
\usepackage[english]{babel} 
\usepackage[T1]{fontenc}
\usepackage{graphicx}
\usepackage{tikz}
\usepackage{numprint}
\usepackage[switch]{lineno}
\usepackage{fancybox}
\usepackage[utf8]{inputenc}
\usepackage{graphicx}
\usepackage{booktabs}
\usepackage[xcolor=orange,todonotes={colorinlistoftodos,prependcaption,textsize=footnotesize}]{changes}
\usepackage{xcolor}
\usepackage[most]{tcolorbox}
\usepackage{fancybox}

\newcommand\ourlogic{HLS\xspace}
\newcommand{\translation}{\textcolor{red}{\ensuremath{\mathfrak{h}}}}
\newcommand{\tracetrans}{\textcolor{blue}{\ensuremath{\mathfrak{t}}}}

\newcommand\resq[1]{
\noindent 
\fcolorbox{green!40!black}{green!5}{\noindent 
 \parbox{0.98\columnwidth}{\noindent  #1}}\\
}

 \definecolor{myBlue}{RGB}{0,133,255}\definecolor{myGreen}{RGB}{0,115,0}

\newcommand\interpretation[1]{\ensuremath{\llbracket #1 \rrbracket}}

\usepackage{makecell}

\npthousandsep{\,}

\usepackage{algorithm}

\usepackage{algorithmicx}
\usepackage{algpseudocode}

\newcommand\real{\mathbb{R}}
\newcommand\signal{\ensuremath{s}}
\newcommand\signals{\ensuremath{S}}

\definecolor{keywordcolor}{RGB}{127,0,85}
\newcommand{\lit}[1]{\textbf{\texttt{\textcolor{keywordcolor}{#1}}}}
\newcommand\synt[1]{\texttt{#1}}
\newcommand{\attt}{\ensuremath{\mathbin{\lit{@t}}}}
\newcommand{\atst}{\ensuremath{\mathbin{\lit{@i}}}}
\newcommand\rqonerequirements{212}
\newcommand\rqoneHLSpercentage{100}
\newcommand\rqoneTempsypercentage{68}
\newcommand\rqoneSTLpercentage{48}

\newcommand\rqoneTempsynumber{145}
\newcommand\rqoneSTLnumber{102}

\newcommand\numsimulationtraces{20}
\newcommand{\minnumberofentries}{41844}
\newcommand{\maxnumberofentries}{1202241}
\newcommand{\averageumberofentries}{389771}
\newcommand{\stdevnumberofentries}{393718}

\newcommand{\minnumberofmem}{1.7}
\newcommand{\maxnumberofmem}{58.9}
\newcommand{\averagenumberofmem}{17.6}
\newcommand{\stdevnumberofmem}{19.4}

\newcommand{\aoneminnumberofentries}{2}
\newcommand{\aonevmaxnumberofentries}{17321}
\newcommand{\aoneaverageumberofentries}{2071}
\newcommand{\aonestdevnumberofentries}{3840}

\newcommand{\aoneminmem}{15}
\newcommand{\aonemaxmem}{5.9}
\newcommand{\aoneavgmem}{0.1}
\newcommand{\aonestdmem}{0.4}

\newcommand{\atwominnumberofentries}{2}
\newcommand{\atwomaxnumberofentries}{2360674}
\newcommand{\atwoaverageumberofentries}{52406}
\newcommand{\atwostdevnumberofentries}{185875}

\newcommand{\atwominmem}{15}
\newcommand{\atwomaxmem}{90.0}
\newcommand{\atwoavgmem}{2.3}
\newcommand{\atwostdmem}{8.4}

\newcommand{\tracerequirementcombinations}{747}
\newcommand{\tracerequirementcombinationsone}{320}
\newcommand{\tracerequirementcombinationstwo}{178}
\newcommand{\tracerequirementcombinationsthree}{249}
\newcommand{\tracerequirementcombinationstwoencodings}{1494}
\newcommand{\totalexperiments}{14940}

\newcommand\aonepositiveperc{53.9}
\newcommand\aonepositiveavg{80.2}
\newcommand\aonepositivemax{2693.0}
\newcommand\aonepositivemin{0.01}
\newcommand\aonepositivedev{334.7}

\newcommand\aonenegativeperc{12.1}
\newcommand\aonenegativeavg{14.2}
\newcommand\aonenegativemax{513.9}
\newcommand\aonenegativemin{0.01}
\newcommand\aonenegativedev{57.9}

\newcommand\aoneundecidedperc{1.6}
\newcommand\aoneundecidedavg{6.5}
\newcommand\aoneundecidedmax{7.4}
\newcommand\aoneundecidedmin{5.8}
\newcommand\aoneundecideddev{0.6}

\newcommand\aonetimeout{0.5}
\newcommand\aonemaxdepth{13.0} 
\newcommand\aonoutofMemory{18.9}

\newcommand\aoneverdictintime{66.0\%}

\newcommand\aonewithintimeoutpercentage{67.6$\%$}

\newcommand{\numexperimentsatwo}{7470}

\newcommand\atwopositiveperc{53.8}
\newcommand\atwopositiveavg{102.5}
\newcommand\atwopositivemax{3432.9}
\newcommand\atwopositivemin{0.01}
\newcommand\atwopositivedev{331.7}

\newcommand\atwonegativeperc{20.7}
\newcommand\atwonegativeavg{96.5}
\newcommand\atwonegativemax{3143.5}
\newcommand\atwonegativemin{0.01}
\newcommand\atwonegativedev{379.8}

\newcommand\percentagedefinitiveatwo{74.5}

\newcommand\atwoundecidedperc{2.2}
\newcommand\atwoundecidedavg{8.7}
\newcommand\atwoundecidedmax{12.3}
\newcommand\atwoundecidedmin{5.4}
\newcommand\atwoundecideddev{2.1}

\newcommand\atwotimeout{23.3}

\newcommand\atwowithintimeoutpercentage{76.7}

\newcommand{\numexperimentsonlyHLS}{337}

\newcommand{\numrunsonlyHLS}{3370}

\newcommand{\onlyHLSverdicpercentage}{67.9\%}

\newcommand{\thedorebreachcomparisontraced}{103}
\newcommand{\thedorebreachcomparisontracedexperiments}{1030}

\newcommand{\Theodorevsbreachpercentage}{95.1}
\newcommand{\Theodorevsbreachavg}{81.4}
\newcommand{\Theodorevsbreachmax}{2506.2}
\newcommand{\Theodorevsbreachmin}{0.01}
\newcommand{\Theodorevsbreachstd}{345.7}

\newcommand{\breachpercentage}{100}
\newcommand{\breachavg}{0.03}
\newcommand{\breachmax}{0.1}
\newcommand{\breachmin}{0.02}
\newcommand{\breachstd}{0.007}

\newcommand{\thedoretempsycomparisontraced}{162}
\newcommand{\thedoretempsycomparisontracedexperiments}{1620}

\newcommand{\Theodorevstempsypercentage}{72.2}
\newcommand{\Theodorevstempsyavg}{69.6}
\newcommand{\Theodorevstempsymax}{2506.2}
\newcommand{\Theodorevstempsymin}{0.01}
\newcommand{\Theodorevstempsystd}{317.6}

\newcommand{\sbtempsypercentage}{94.1}
\newcommand{\sbtempsyavg}{30.1}
\newcommand{\sbtempsymax}{3440.0}
\newcommand{\sbtempsymin}{0.09}
\newcommand{\sbtempsystd}{310.1}

\begin{document}

\title{
Trace-Checking CPS Properties: \\
Bridging the Cyber-Physical Gap 
}

\IEEEoverridecommandlockouts

\author{
\IEEEauthorblockN{Claudio Menghi}
\IEEEauthorblockA{
University of Luxembourg\\
Luxembourg, Luxembourg\\
claudio.menghi@uni.lu}
\and
\IEEEauthorblockN{Enrico Vigan\`o \IEEEauthorrefmark{1}}
\IEEEauthorblockA{
University of Luxembourg\\
Luxembourg, Luxembourg\\
enrico.vigano@uni.lu
\thanks{\IEEEauthorrefmark{1}This work was carried out while the
  author was a student at Politecnico di Milano (Italy), during an
  internship at the SnT Centre of the University of Luxembourg.}
}
\and
\IEEEauthorblockN{Domenico Bianculli}
\IEEEauthorblockA{
University of Luxembourg\\
Luxembourg, Luxembourg\\
domenico.bianculli@uni.lu}
\and
\IEEEauthorblockN{Lionel~C. Briand}
\IEEEauthorblockA{
University of Luxembourg\\
Luxembourg, Luxembourg\\
University of Ottawa\\
Ottawa, Canada \\
lionel.briand@uni.lu}
}

\maketitle    
\thispagestyle{plain}
\pagestyle{plain}

\begin{abstract}
Cyber-physical systems combine software and physical components. 
Specification-driven trace-checking tools for CPS usually provide users with a specification language to express the requirements of interest, and an automatic procedure to check whether these requirements hold on the execution traces of a CPS.  Although there exist several specification languages for CPS,  they are often not sufficiently expressive to allow the specification of complex CPS properties related to the software and the physical components and their interactions.

In this paper, we propose (i) the Hybrid Logic of Signals (HLS), a logic-based language that allows the specification of complex CPS requirements, and (ii)  ThEodorE, an efficient SMT-based trace-checking procedure. This procedure reduces the problem of checking a CPS requirement over an execution trace, to checking the satisfiability of an SMT formula.

We evaluated our contributions by 
using a representative industrial case study in the satellite domain.
We assessed the expressiveness of HLS by considering \rqonerequirements\ requirements of our case study.
HLS could express all the \rqonerequirements\ requirements. 
We  also assessed the applicability of ThEodorE by running the trace-checking procedure for \tracerequirementcombinations\  trace-requirement combinations.
ThEodorE was able to produce a verdict in \percentagedefinitiveatwo\% of the cases.
Finally, we compared HLS and ThEodorE with other specification languages and trace-checking tools from the literature.
Our results show that, from a practical standpoint, our approach offers a better trade-off between expressiveness and performance.  \end{abstract}

\section{Introduction}
\label{sec:intro}
Cyber-physical systems (CPSs) combine cyber and physical capabilities~\cite{platzer2018logical}.
Cyber capabilities are typically provided by software components that sense and act on the physical environment, while
physical capabilities are provided by the environment in which the software is deployed. 
Therefore, CPSs combine  software and  physical dynamics.
Physical dynamics are typically modeled through formalisms that capture the  continuous evolution---according to physical laws---of the  environment over time (e.g., differential equations); the corresponding behaviors are typically represented as continuous signals.
Software (i.e., cyber) dynamics are typically modeled with discrete event systems (e.g., finite state machines), whose behavior is typically represented by a sequence of events.
Cyber-physical systems exhibit \emph{hybrid} dynamics since they combine both physical and software capabilities.

Engineers collect traces (i.e., logs) describing the behavior of a CPS both when the CPS is simulated and,  by means of instrumentation and logging mechanisms, during the actual execution of the CPS.
A trace is a sequence of records that contain some information about the execution (or the simulation) of the  cyber-physical components  (e.g., the state of the system variables).
Trace records are usually labeled with time-stamps representing the time instants at
which the recorded information was obtained. 

Engineers  analyze these traces to check whether they conform to the  system's requirements specifications; this activity can be automated by means of trace-checking tools.
Specification-driven trace-checking tools usually take as input a trace to be analyzed and a requirement specification; they yield a Boolean verdict indicating whether the trace satisfies the specification. The algorithms implemented by trace-checking tools are typically language-specific.
In the context of trace checking, there exist two main categories of languages used for specifying  CPS requirements: \emph{time-based} and \emph{sequence-based} languages.

Time-based languages (e.g., STL~\cite{maler2004monitoring}, 
STL$^\ast$~\cite{brim2014stl},
RFOL~\cite{FSE2019}, 
SFO~\cite{bakhirkin2018first},
TPTL~\cite{alur1993real}.
and \tempsydslname ~\cite{ase2020})  interpret the records of the cyber and  physical components as  signals over a  \emph{time domain}. 
Specifications written in a time-based language express time relations over the occurrence of events.
Such languages are suitable to express CPS requirements related to physical quantities; an example of such requirement is P1: ``between \SI{2}{\s} and \SI{10}{\s} (measured starting from the origin of the trace) the speed of the satellite is lower than  \SI{10}{\m/\s}''. 
However, time-based languages are not easily amenable to specifying requirements related to software components. As an example, let us consider the requirement P2: ``\emph{whenever the satellite changes its mode from safe to normal}, the speed of the satellite decreases''. 
To express the first part of this requirement (marked in italics), one should specify that 1) in the trace there are  two \emph{consecutive} records;  2) the first record captures that the satellite is in ``safe mode''; and 3) the second record captures that the satellite is in ``normal mode''. This requirement cannot be easily expressed in time-based languages since they generally do not provide constructs specifically designed to specify the first condition, i.e., that a record immediately follows another one in the trace. Indeed, expressing such a condition requires the specification language to provide access to the indices (i.e., positions in the trace) of the different records.

On the other hand, in sequence-based languages---such as LTL~\cite{emerson1986sometimes} (and domain-specific languages based on one of its extensions, like the one in the SpeAR tool~\cite{fifarek2017spear}), \textsc{fretish}~\cite{giannakopoulou2020formal}, and CoCoSpec~\cite{champion2016cocospec}---traces are sequences of consecutive records, whose temporal model is represented by the sequence of \emph{discrete indices} of the records.
This class of languages interprets the records of the CPS software and physical components as discrete-time signals.
Specifications in these languages constrain the indices in which events can  occur; such specifications are used to express properties that mostly refer to the CPS software components, such as the first part of the aforementioned P2 property.
However, these languages cannot express time relations over the occurrence of events, such as the one in property P1.

A third class of specification languages is the one of \emph{hybrid} languages (e.g.,
STL-MX~\cite{STLmx},  HyLTL~\cite{bresolin2013hyltl}, HRELTL~\cite{CIMATTI201554}, Differential Dynamic Logic~\cite{platzer2008differential}, 
 HTL~\cite{henzinger1992towards}), which support the specification of both continuous and discrete behaviors. 
However, these languages  typically extend existing languages (e.g., LTL) to support the specification of hybrid behaviors in specific contexts (e.g., using signal derivatives). 
Therefore, they provide ad-hoc solutions that inherit some of the intrinsic limitations of the base language, thus hindering the expressiveness of the resulting hybrid language. For example, a hybrid language based on LTL cannot support metric operators to constrain the time distance between events.
 
The goal of this paper is to tackle the challenge of specifying hybrid behaviors of CPSs, in a way amenable to practical and efficient trace-checking. 
To reach this goal we propose:
\begin{asparaenum}[(i)]
    \item the \emph{Hybrid Logic of Signals (\ourlogic)}, a new specification language tailored to specifying CPS requirements. \ourlogic allows engineers to express CPS requirements as properties (i.e., specifications) that refer both to the time-stamps and to the indices of the records of CPS traces. In this way, \ourlogic specifications can easily express the behavior of  both cyber and physical components, as well as their interactions.
    \item \emph{\app}, an efficient trace-checking approach for properties expressed in \ourlogic. 
\app reduces the problem of checking an \ourlogic property on a trace to a
satisfiability problem, which can be solved  using off-the-shelf
Satisfiability Modulo Theories (SMT) solvers. 
The latter have efficient decision procedures for several background theories, 
thus making it possible to check whether a formula expressed in a first-order logic is satisfiable.
\end{asparaenum}

We evaluated our contribution using an \emph{industrial} case study in the satellite domain, in collaboration with the engineers who developed the satellite's on-board system.
\begin{asparaenum}[(i)]
      \item  We assessed the expressiveness of \ourlogic by checking whether  it could express the \rqonerequirements\  requirements of our case study. Our results show that \ourlogic could express all the requirements of our case study. We also compared \ourlogic with SB-TemPsy-DSL~\cite{ase2020} and STL~\cite{maler2004monitoring}, two specification languages proposed in the literature and for which trace-checking tools are available. The results show that \ourlogic\ is significantly more expressive than SB-TemPsy-DSL and STL, which could only express \rqoneTempsynumber\ and \rqoneSTLnumber\ requirements, respectively.
\item We evaluated the trace-checking support provided by \app by assessing its applicability on 20 large traces provided by our industrial partner and obtained  by  simulating  the  behavior  of  the  satellite across representative,  different  scenarios. We ran the \app trace-checker on \tracerequirementcombinations\ trace-requirement combinations.
\app completed the verification in \percentagedefinitiveatwo\% of the cases within one hour,  a reasonable time-out considering typical CPS development contexts.
 \app  yielded a verdict for \onlyHLSverdicpercentage\ of the  \numexperimentsonlyHLS\ trace-requirement combinations  containing a requirement that can  not be verified by any of the other trace-checkers.
 We compared the applicability of \app with SB-TemPsy-Check~\cite{ase2020} and Breach~\cite{DBLP:conf/cav/DonzeFM13}, for the trace-requirement combinations containing requirements  expressible in  SB-TemPsy-DSL and STL. 
For these combinations, SB-TemPsy-Check and Breach were 21.9\% and 4.9\% more often applicable than \app, respectively. 
SB-TemPsy-Check and Breach were also more efficient,
but  not  to  a  point  where  it  had practical implications. 
\end{asparaenum}
Our results show that \app is broadly applicable as it allows engineers to specify a large variety of requirements while providing an efficient trace-checking procedure. Since in practical applications it is generally difficult to know in advance which requirement types engineers will need to specify, 
our findings suggest that  \app  is good default choice. 
However, if \app is not able to produce a verdict, and the requirement are expressible in SB-TemPsy-DSL or STL, engineers should then use SB-TemPsy-Check or Breach.

The paper is organized as follows. 
Section~\ref{sec:casestudy} describes our case study.
Section~\ref{sec:hls}  illustrates the  syntax and semantics of  \ourlogic. 
Section~\ref{sec:Theodore} presents \app.
Section~\ref{sec:evaluation} evaluates our contribution based on an industrial case study.
Section~\ref{sec:related} discussed related work. Section~\ref{sec:conclusion} concludes the paper.

\section{Case Study and Motivations}
\label{sec:casestudy}
Our industrial partner LuxSpace~\cite{Luxspace} developed, in collaboration with ESA~\cite{esa} and ExactEarth~\cite{exactEarth}, a maritime satellite to collect tracking information from vessels operating 
on Earth and to relay those data to the ground.
This is a representative CPS made of complex software component interacting with many actuators and sensors and the physical environment where the satellite is to be deployed. 
This system should satisfy many varied requirements regarding the behavior of the software system itself but also its interactions with hardware and the satellite physical dynamics in space. 
Its development relies on technologies and practices typically seen in CPS contexts, e.g., Model-in-the-loop development with Simulink\textsuperscript{\tiny\textregistered}.

Software engineers check the compliance of the satellite behavior to its requirements~\cite{SatellitePhases} both while the software is being developed and at run time. This is done by
\begin{enumerate*}[(i)]
    \item collecting execution traces of the system, and
    \item checking whether those traces satisfy the system requirements.
\end{enumerate*} 

Figure~\ref{fig:traceExample} shows a fragment of an execution trace, which we will use to motivate this work. 
A trace is a sequence of records that contain some information about the execution  of the system.
In this example, the records include data about the \emph{angular rate} (\texttt{ang-rate}) and the \emph{(satellite) mode} (\texttt{mode}). 
The angular rate is a physical quantity represented by a real value  measured by sensors.
The mode is an enumeration of values that represent the state of the satellite software.
There are four different modes: ``Idle Mode'', ``Safe Spin Mode'', ``Normal Mode Coarse'', and ``Normal Mode Fine'', which are represented in the trace by the values $0$, $1$, $2$, and $3$, respectively. 
In addition, each record is associated with a timestamp,  representing  the  time  instant  at  which  the recorded information was obtained, and a progressive index value.

\begin{figure}
\begin{tikzpicture}
	\pgfmathsetmacro{\ilocationangularrate}{0.25}
	\pgfmathsetmacro{\ilocationmode}{-0.25}
	\pgfmathsetmacro{\ilocationtimestamp}{-0.75}
	\pgfmathsetmacro{\ilocationindex}{-1.2}
	\draw[dashed] (-2,0.5) -- (6.8,0.5);
		\draw node at (-1,\ilocationangularrate) {\footnotesize \texttt{ang-rate}};
	\draw node at (0.5,\ilocationangularrate) {\small $20.1$};
	\draw node at (1.5,\ilocationangularrate) {\small $22.2$};
	\draw node at (2.5,\ilocationangularrate) {\small $23.3$};
	\draw node at (3.5,\ilocationangularrate) {\small $20.4$};
	\draw node at (4.5,\ilocationangularrate) {\small $21.1$};
	\draw node at (5.5,\ilocationangularrate) {\small $3.2$};
	\draw node at (6.5,\ilocationangularrate) {\small $1.1$};
	\draw[dashed] (-2,-0) -- (6.8,-0);
		\draw node at (-1,\ilocationmode) {\footnotesize \texttt{mode}};
	\draw node at (0.5,\ilocationmode) {\small $0$};
	\draw node at (1.5,\ilocationmode) {\small $1$};
	\draw node at (2.5,\ilocationmode) {\small $0$};
	\draw node at (3.5,\ilocationmode) {\small $0$};
	\draw node at (4.5,\ilocationmode) {\small $3$};
	\draw node at (5.5,\ilocationmode) {\small $3$};
	\draw node at (6.5,\ilocationmode) {\small $3$};
		\draw[dashed] (-2,-0.5) -- (6.8,-0.5);
		\draw node at (-1,\ilocationtimestamp) {\small timestamp};
	\draw node at (0.5,\ilocationtimestamp) {\small $0$};
	\draw node at (1.5,\ilocationtimestamp) {\small $0.2$};
	\draw node at (2.5,\ilocationtimestamp) {\small $0.9$};
	\draw node at (3.5,\ilocationtimestamp) {\small $1.8$};
	\draw node at (4.5,\ilocationtimestamp) {\small $3.0$};
	\draw node at (5.5,\ilocationtimestamp) {\small $4.9$};
	\draw node at (6.5,\ilocationtimestamp) {\small $5.7$};
	\draw[dashed] (-2,-1) -- (6.8,-1);
	\draw node at (-1,\ilocationindex) 
	{\small index};
	\draw node at (0.5,\ilocationindex) {\small $0$};
	\draw node at (1.5,\ilocationindex) {\small $1$};
	\draw node at (2.5,\ilocationindex) {\small $2$};
	\draw node at (3.5,\ilocationindex) {\small $3$};
	\draw node at (4.5,\ilocationindex) {\small $4$};
	\draw node at (5.5,\ilocationindex) {\small $5$};
	\draw node at (6.5,\ilocationindex) {\small $6$};
    \draw[dashed] (-2,-1.5) -- (6.8,-1.5);
	\pgfmathsetmacro{\llocation}{-1.5}
	\pgfmathsetmacro{\dlocation}{-2}
	\pgfmathsetmacro{\xlocation}{-2.5}
	\pgfmathsetmacro{\edgelocation}{-3.5} ;
	\draw (3.1,0.6) -- (3.9,0.6) -- (3.9,-1.6) -- (3.1,-1.6) -- (3.1,0.6);
		\draw node at (3.9,-1.8) {\small Record $r_3$};
	\end{tikzpicture} \caption{A fragment of an execution trace of our case study.}
\label{fig:traceExample}
\end{figure} 

The requirements to be checked on the system traces  refer both to the software and to the physical  dynamics of the satellite.  For example, let us consider requirement $\mathcal{R}1$:
\emph{
Whenever \textbf{the satellite mode switches from ``Idle Mode'' to ``Normal Mode Fine''}, 
the satellite angular rate shall reach a value lower than \SI{1.5}{\degree\per\s} \textbf{within \SI{10}{\s}}.  
Moreover, the angular rate shall  stabilize around \textbf{an arbitrary  value $c$} lower than or equal to \SI{1.5}{\degree\per\s}.
}
$\mathcal{R}1$ specifies a constraint on a physical quantity, i.e., the angular rate of the satellite, which shall be ensured as a reaction to a software change, i.e., the satellite switching its mode from ``Idle'' to ``Normal Mode Fine''. 

 One way to express that \emph{\textbf{the mode of the satellite switches from ``Idle Mode'' to ``Normal Mode Fine''}},   is  to specify that the trace contains:
 \begin{enumerate*}
     \item  two  records with \emph{consecutive} indices;
     \item the first record captures that the satellite is in ``Idle Mode'';
     \item the second record captures that the satellite is in ``Normal Mode Fine''.
 \end{enumerate*}
 This requirement cannot be easily expressed in the vast majority of time-based languages since they do not provide  access  to  the indices  of the different records.
 To compensate for this limitation when using time-based languages, engineers can apply ad-hoc solutions, 
 such as adding a new Boolean flag to the trace records. In our example,  such a flag would be true whenever the mode of the satellite switches from ``Idle Mode'' to ``Normal Mode Fine''. 
 In this way, the aforementioned requirement fragment would be rephrased as \emph{\textbf{the flag \texttt{switch-from-IDLE-to-NORMAL-MODE-FINE} is true}}.
 However, this is impractical in real scenarios because:
\begin{enumerate*}[(i)]
\item the number of flags to add in the trace records can quickly grow and become unmanageable. 
 For example, given the four possible values for the satellite mode in our case study, to consider all possible combinations for switching satellite mode, engineers would need to add 16 values in each record (one for each mode switching combination).
 \item The requirement is reformulated and its connection to the actual software component behavior is lost. 
 \end{enumerate*}

Furthermore, requirement $\mathcal{R}1$ cannot be expressed using sequence-based languages because they do not support time relations over the occurrence of events. 
More specifically, expressing that ``the [\dots] angular rate shall reach [\dots] \textbf{within \SI{10}{\s}}''
requires to access the timestamps associated with the trace records (and compute a distance). This feature is not provided by sequence-based languages.

Moreover, to the best of our our knowledge, among the time-based and sequence-based languages mentioned in the previous section, 
SFO~\cite{bakhirkin2018first} is the only  language that  allows users to use quantified variables in specifications, 
(as in ``(there exist)  \textbf{an arbitrary  value} $c$ lower than or equal to \SI{1.5}{\degree\per\s} around which [\dots] shall  stabilize''.
This type of requirements is quite  common in practical CPS applications, since engineers often want to check that the system stabilizes around a given value (e.g., the steady-state value).
Although engineers know some properties of the steady-state value $c$ (i.e., $c$ shall be lower than or equal to  \SI{1.5}{\degree\per\s}), 
they generally do not know its exact value, which has to be indicated as a generic variable in the requirement specification. 

This example, extracted from our case study, shows the need for an expressive language for specifying hybrid behaviors of CPSs.
In the next section, we will introduce a new specification language for CPSs, which overcomes the limitations---in terms of expressiveness---of state-of-the-art languages  and is supported by an effective trace-checking procedure.

\section{Hybrid Logic of Signals}
\label{sec:hls}
In this section, we illustrate \emph{\ourlogic (Hybrid Logic of Signals)}, our new specification language for CPSs.
We first discuss the design goals of the language  (section~\ref{sec:bkg:principles}).
Then, we define the mathematical model of the traces considered in this work (section~\ref{sec:bkg:traces}). 
Finally, we present the syntax  (section~\ref{sec:syntax}) and the semantics (section~\ref{sec:semantics}) of the language. 

\subsection{Design goals}
\label{sec:bkg:principles}
We designed \ourlogic to provide a language for specifying CPS properties that seamlessly combine the features of sequence-based and time-based languages. 
Therefore, \ourlogic extends existing time-based languages (e.g., STL~\cite{maler2004monitoring}, RFOL~\cite{FSE2019}, and SFO~\cite{bakhirkin2018first}) 
and sequence-based languages (e.g., LTL~\cite{emerson1986sometimes}, \textsc{fretish}~\cite{giannakopoulou2020formal}, and CoCoSpec~\cite{champion2016cocospec}) 
to allow engineers \emph{to refer both to trace indices and to timestamps} in the logical specifications, \emph{to arbitrarily combine them} to define properties describing the expected behavior of a CPS, and to express properties by quantifying over the values of the variables.
More specifically, \ourlogic allows engineers to use first-order \emph{existential} and \emph{universal} quantifiers with:
\begin{itemize}
    \item \emph{timestamp variables}, to express properties that refer to specific time instants and to the distance among them,
     such as ``\emph{there exists a time instant $t$ within \SI{10}{\s} from the current time instant [\dots]}'';
    \item  \emph{(trace) index variables}, to express properties that refer to the indices of trace records,
    such as ``\emph{for every trace index~$i$, such that the  corresponding record captures that the satellite is in ``Idle Mode'', and the immediately following record (at trace index $i+1$) captures that the satellite is in ``Normal Mode Fine'' [\dots]}'';
    \item \emph{real-valued variables}, to express properties that refer to arbitrary signal values,
    such as ``\emph{there exists a value $c$ lower than or equal to \SI{1.5}{\degree \per \s} around which the signal \emph{ang-rate} shall stabilize}''. 
\end{itemize} 
Additionally, \ourlogic supports specifications that use:
\begin{itemize}
    \item the value of a signal  \emph{at a certain timestamp} or associated with a record \emph{at a certain index};
     \item  the timestamp associated with the record  \emph{at a certain index};
    \item the index of the record \emph{with a certain timestamp};
    \item expressions combining  time variables, trace indices, and real-valued variables, using arithmetic  and relational  operators.
\end{itemize}

\subsection{Traces}
\label{sec:bkg:traces}
Let $\mathbb{J}=\{0, 1, 2, \ldots, j, \ldots,  m\}$, with elements $j \in \mathbb{N}$, be a set of indices. 
Let $\mathbb{T}$ be an  interval of $\real$; we call $\mathbb{T}$ a time domain. 
Let $\signals=\{ \signal_1, \signal_2, \ldots , \signal_i, \ldots, \signal_n \}$ be a set of  variables  (hereafter called  ``signals'') of the systems being monitored, with $s_i \in \real$.
A trace $\pi$ is a finite sequence of records $r_0,r_1,\ldots ,r_j, \ldots, r_m$, with $j \in \mathbb{J}$.

Each record $r_j$ is a tuple $\langle j, t, v_{1}, v_{2}, \ldots , v_{n} \rangle$, 
where $j \in \mathbb{J}$ is the index associated with the record,  $t \in \mathbb{T}$ is the timestamp at which the recorded information was obtained,
and  $ v_{1}, v_{2}, \ldots, v_{n} \in \real$ are the values associated with signals $\signal_1, \signal_2, \ldots , \signal_n$ in the record.
For a trace $\pi$ we use the array notation ``$[j]$'' to denote the
$j$-th record of $\pi$, and we use the dot notation to denote an
element of a record; we also introduce the  notation $t_j$, short for $\pi[j].t$ for a given trace $\pi$.
For example, let $\pi^e$ be the fragment of the trace depicted in
Figure~\ref{fig:traceExample}; it contains seven records.
Record $r_3$ is denoted by $\pi^e[3]$; it is represented by the tuple
$\langle 3, 1.8, 0, 20.4 \rangle$, where $\pi^e[3].t=t_3=1.8$ is the value
of the timestamp, $\pi^e[3].\texttt{mode}=0$ is the value of signal
\texttt{mode}, and $\pi^e[3].\texttt{ang-rate}=20.4$ is the value of signal \texttt{ang-rate}.

We assume that the values associated with the timestamps are monotonically increasing, i.e., $t_j<t_{j+1}$, since records refer to consecutive timestamps. 
We say that a trace has a \emph{fixed sample rate} $\mathit{sr}$ if,  for every $j, 0\leq j <m$,  $t_{j+1}-t_j=\mathit{sr}$, 
where $\mathit{sr}$ is a constant value; otherwise, we say that the trace has a \emph{variable sample rate}. 
For example, trace $\pi^e$ in Figure~\ref{fig:traceExample} has a variable sample rate.

Additionally, we introduce a function $\iota_{\pi} \colon  \mathbb{T} \to \mathbb{J}$: 
given a timestamp value $t$, $\iota_{\pi}(t)$ is the value of the index $j$ of the record in $\pi$ 
with the highest  timestamp $t_j$ such that $t_j<=t$; we will omit the
trace subscript when it is clear from the context. For example, for trace $\pi^e$ in Figure~\ref{fig:traceExample},
$\iota_{\pi^e}(2.5)=3$. 
In this work, we consider two definitions of $\iota$:
\begin{align*}
\iota^{V}(t) &::=  [t_0\leq t] \cdot \left[t < t_1\right] \cdot 0 + 
[t_1\leq t] \cdot [t < t_2] \cdot 1 +{}
 \nonumber \\
& \ldots  {}+[t_{m-1}\leq t] \cdot [t < t_m] \cdot (m-1) +
  [t_{m}=t] \cdot m \\
    \iota^{F}(t) &::= \left\lfloor \frac{t}{\mathit{sr}} \right\rfloor  
\end{align*}

Definition $\iota^{V}(t)$  assumes that the
trace has a variable sample rate. Notice that the notation $[P]$,
where $P$ is a logical predicate, is the Iverson bracket; it evaluates
to 1 if $P$ is true, and to 0 otherwise. The resulting arithmetic
formula checks where the timestamp $t$ provided in input is situated
w.r.t. the timestamps of the trace (i.e., $t_0, t_1, \ldots, t_m$), and
returns the value of the index of the record that has the highest
timestamp that is smaller than or equal to $t$.  For example, if the
parameter $t$ is greater than timestamp $t_2$ and lower than timestamp
$t_3$, the only expression in $\iota^{V}(t)$ that does not evaluate to 0 is
$[t_2\leq t] \cdot [t < t_3] \cdot 2$; therefore the index returned
will be~2.

Definition $\iota^{F}(t)$  assumes that
the trace has a fixed sample rate. In such as case, the index
associated with a timestamp can be simply retrieved by computing the
floor of the ratio of the timestamp $t$ over the sample rate
$\mathit{sr}$.

In this work, we assume that all the variables are sampled at each timestamp.
This is a necessary requirement to enable the evaluation of the satisfaction of the system requirements at each timestamp. 
For systems that do not sample all the variables at each timestamp, engineers can use a pre-processing step 
to generate values to be assigned to variables for which the value is missing at  certain timestamps. In this work, we consider two complementary
pre-processing strategies:
 \begin{asparaenum}[{$\mathcal{A}$}1:]
\item In each record, an interpolation function (e.g., piece-wise constant,  linear, cubic) specific to each signal, is used to generate values for unassigned variables. 
Notice that this approach does not alter the original sample rate of the trace, since it keeps the same records as the original trace and only generates (in each record) values for the \emph{unassigned} variables.
\item If the trace has a variable sample rate, it is  converted into a trace with a fixed sample rate.  This is done by generating a fresh set of records with a fixed sample rate
 equal to the smallest sample rate (i.e., the 
minimum time distance between two records) of the original trace, and by using the interpolation functions (as in the case of strategy $\mathcal{A}1$)  to generate the values of \emph{all} variables.
\end{asparaenum}
As we will discuss in Section~\ref{sec:evaluation}, the strategy used to generate the values of unassigned variables determines the trace accuracy. The latter influences the trace checking verdict and may  impact on the correctness  of the trace-checking procedure.

\subsection{Syntax}
\label{sec:syntax}

An \ourlogic formula  is defined according to the grammar presented in \figurename~\ref{tab:grammar}, whose start symbol is \texttt{p}. 
In the grammar, we use the symbol $f$ to represent a generic (binary) arithmetic function; the symbol $\mid$ separates alternatives.
In the following, we illustrate the various language constructs; in the explanations, we will refer to
the set  $\mathit{TV}=\{\tau_0,\tau_1, \dots\}$ of timestamp variables over $\mathbb{T}$, 
the set $\mathit{IV}=\{\sigma_0,\sigma_1, \dots\}$ of index variables over $\mathbb{J}$,
and the set $\mathit{RV}=\{\rho_0, \rho_1, \dots\}$ of real-valued variables over $\real$.

\begin{asparadesc}
\item A \emph{term} (non-terminal \synt{tm}) can be either a \emph{time term}, an \emph{index term}, or a \emph{value term}.
\item  A \emph{time term} (non-terminal \synt{tt})
allows engineers to refer to timestamps in the specifications.
A time term can be a timestamp variable $\tau \in \mathit{TV}$, a literal denoting a value $t \in \mathbb{T}$,
the value returned by the operator \lit{i2t}, or an arithmetic expression over these entities.
The  operator \lit{i2t}(\synt{it}) takes an index term as argument and returns  the timestamp associated with the record at the (trace) index \synt{it}.
An example of time term is the expression $\tau_0+5.5+\lit{i2t}(2)$.
\item An  \emph{index term} (non-terminal \synt{it}) allows engineers to refer to trace indices in the specifications.
An index term can be an index variable $\sigma \in \mathit{IV}$, a literal denoting a value $j \in \mathbb{J}$, 
the value returned by the operator \lit{t2i}, or an arithmetic expression over these entities.
The operator \lit{t2i}(\synt{tt}) takes a time term as argument and 
returns the index $j$ of the trace record with timestamp  $t_j$, 
where  $t_j$ is the highest timestamp value for which  $t_j \leq \synt{tt}$. 
An example of index term is the expression $\sigma_0+2+\lit{t2i}(3.3)$.
\item  A \emph{value term} (non-terminal \texttt{vt}) allows engineers to refer to real values (e.g., signal values) in the specifications.
A value term can be a real-valued variable $\rho \in \mathit{RV}$, a literal denoting a value $x \in \real$,
the value of a signal returned by the operators \atst\ (``at index'')  and \attt\ (``at timestamp''), or an arithmetic expression over these entities.
The \atst\ operator is an infix operator that takes two arguments: a signal $s$ and an index term \synt{it}; 
it returns the value of  signal $s$ associated with the record at the (trace) index  \synt{it}.
Similarly, the \attt\ operator is an infix operator that takes two arguments: a signal $s$ and a time term \synt{tt}; 
it returns the value of  signal $s$ associated with a record at timestamp $t_j$, where $t_j$ is the highest timestamp value in the trace for which   $t_j \leq \synt{tt}$.
An example of value term is the expression $(s_1 \atst 2) + (s_2 \attt 3.3)+\rho_0+5.2$, where $s_1$ and $s_2$ are signals, $2$ is an index term, $3.3$ is a time term, $\rho_0$ is a real-valued variable, and $5.2$ is a numeric literal.
\item A \emph{formula} (non-terminal \synt{p}) is a relational expression over terms,
a logical expression over other formulae defined using  Boolean connectives, 
or an existentially quantified formula. 
As anticipated in section~\ref{sec:bkg:principles}, \ourlogic supports three types of quantification:
\begin{asparaenum}[(i)]
      \item over timestamp variables, as in  ``\lit{exists} $\tau$ \lit{in} $I_T$ [\dots]'', where $I_T$ is a time range with bounds in $\mathbb{T}$;
      \item over index variables,   as in  ``\lit{exists} $\sigma$ \lit{in} $I_J$ [\dots]'', where $I_J$ is a range of index values with bounds in $\mathbb{J}$;
      \item over real-valued variables, as in  ``\lit{exists} $\rho$ [\dots]''.
\end{asparaenum}
For example, the formula \lit{exists} $\sigma_0$ \lit{in}  $[3,5]$
\lit{such that} $(s_1 \atst \sigma_0) <2.5$ specifies that there exists
a record with index greater than or equal to $3$ and lower than or
equal to $5$, in which the value of signal $s_1$ is less than $2.5$.
\end{asparadesc}

The language is further extended with additional relational operators, 
additional logical connectives (e.g.,  implication (\lit{implies}), conjunction (\lit{and})),
and universal quantifiers (\lit{forall}) on  timestamp variables, index variables, and real-valued variables, using the standard logical conventions.

We now present an application of \ourlogic for the specification of one of the requirements in our case study.
Let us consider a fragment of requirement $\mathcal{R}1$:
\emph{
Whenever the satellite mode switches from ``Idle Mode'' to ``Normal Mode Fine'', 
the satellite angular rate shall reach a value lower than \SI{1.5}{\degree\per\s} within \SI{10}{\s}}. 
We recall that the satellite mode is represented by the signal \texttt{mode}, for which value 0 corresponds to ``Idle Mode'' and value 3 corresponds to ``Normal Mode Fine''; 
also, the angular rate is represented by the signal \texttt{ang-rate}. 
This fragment can  be specified in \ourlogic as: 
\begin{align*}
\lit{f}&\lit{orall}\  \sigma_0\ \lit{in}\  [0,5]\ \lit{such that}   \\ 
&(      (\texttt{mode} \atst \sigma_0)=0\ \lit{and}\ (\texttt{mode} \atst (\sigma_0+1))=3) \\  
&  \lit{implies}\  \lit{exists}\  \tau_0\ \lit{in}\  [\SI{0}{\s},\SI{10}{\s}]\ \lit{such that} \\
& (\texttt{ang-rate} \attt (\tau_0+\lit{i2t}(\sigma_0)) <1.5) ) 
\end{align*}
The sub-formula $((\texttt{mode} \atst \sigma_0)=0\ \lit{and}$ $ (\texttt{mode} \atst (\sigma_0+1))=3)$ 
detects when the satellite switches from ``Idle Mode'' to ``Normal Mode Fine'' over two consecutive records (notice the use of the ``at index'' operator to refer to the consecutive indices $\sigma_0$ and $\sigma_0 + 1$).
This expression is within the scope of the outer universal quantifier, which iterates over
 a range of values for the index variable $\sigma_0$.
 This range  depends on the length of the trace and on the use of $\sigma_0$ in the formula.
 In this case, since the requirement says ``\textit{whenever} [the satellite mode switches\dots]'', 
 in the specification we want to cover the full length of  the trace
 fragment $\pi^e$ in Figure~\ref{fig:traceExample}, where record index values span from 0 to 6.
 We achieve this by setting the  lower bound to zero and the upper bound to  five; 
 in this way, the term $\texttt{mode} \atst (\sigma_0 +1) $ always refers to a record index of the example trace.

The inner quantification over the timestamp variable $\tau_0$ checks whether the angular rate of the satellite reaches a value lower than \SI{1.5}{\degree\per\s} within \SI{10}{\s}.
More specifically, the expression $( \texttt{ang-rate}\, \attt (\tau_0+\lit{i2t}(\sigma_0)) <1.5)$
 represents the value of signal $\texttt{ang-rate}$ at timestamp $\tau_0+\lit{i2t}(\sigma_0)$, where $\tau_0$ is in the interval $[\SI{0}{\s},\SI{10}{\s}]$ (corresponding to the distance of \SI{10}{\s}) and $\lit{i2t}(\sigma_0)$ is the timestamp at which the satellite switches from ``Idle Mode'' to ``Normal Mode Fine'', i.e., the timestamp associated with the record at index $\sigma_0$.

\begin{figure}
\centering
\footnotesize
\begin{tabular}{l@{\ }l@{\ }p{55mm}}
\toprule
\emph{Term} & $\synt{tm}   \Coloneqq $ & $ \synt{tt} \mid \synt{vt}  \mid \synt{it}$ \\
\midrule
\emph{Time Term} & $\synt{tt}  \Coloneqq$ & $\tau \mid t \mid \lit{i2t}(\synt{it})  \mid f(\synt{tt}_1, \synt{tt}_2)$   \\
\midrule
\emph{Index Term} & $\synt{it}  \Coloneqq$ & $\sigma \mid j  \mid 
\lit{t2i}(\synt{tt}) \mid f(\synt{it}_1, \synt{it}_2)$ \\
\midrule
\emph{Value Term} & $\synt{vt}   \Coloneqq$ & $  \rho \mid  x \mid (s \atst \synt{it}) \mid (s \attt \synt{tt})  \mid f(\synt{vt}_1, \synt{vt}_2)$\\
\midrule
\emph{Formula} &$ \synt{p}  \Coloneqq$ & $ \synt{tm}_1 < \synt{tm}_2  \mid \lit{not}\ \synt{p} \mid \synt{p}_1\ \lit{or}\ \synt{p}_2  $\\
 & & $\mid $ $\lit{exists}\ \tau\ \lit{in}\ I_T\ \lit{such that}\  \synt{p}$ \\  & &$\mid $ $\lit{exists}\ \sigma\ \lit{in}\ I_J \lit{such that}\ \synt{p}  $ \\
 & &  $\mid \lit{exists}\ \rho\ \lit{such that}\ \synt{p} $ \\
\bottomrule\\
\end{tabular}
$t \in \mathbb{T},  j \in \mathbb{J}, x \in \real, \tau  \in
\mathit{TV}, \sigma \in \mathit{SV}, \rho \in \mathit{RV}, s \in S$

 \caption{Syntax of the Hybrid Logic of Signals.}
\label{tab:grammar}
\end{figure}

\subsection{Semantics}
\label{sec:semantics}

To evaluate whether an \ourlogic formula is true or false over a trace
$\pi$, we must first define how time, index, and value terms are
interpreted and evaluated.

Let $\mu^{\mathit{TV}}, \mu^{\mathit{IV}}, \mu^{\mathit{RV}}$ be variable assignments,
respectively, for timestamp, index, and real-valued variables; for
example, $\mu^{\mathit{TV}}$ is a mapping from a timestamp variable in
$\mathit{TV}$ to a value in $\mathbb{T}$.  Let $\mu$ denote,
collectively, the family of variable assignment functions
$\mu^{\mathit{TV}}, \mu^{\mathit{IV}}, \mu^{\mathit{RV}}$.
We evaluate a generic term \synt{tm} on a trace $\pi$, using the
variable assignment functions in $\mu$, by means of an interpretation
function $\interpretation{\synt{tm}}_{\pi,\mu}$.

The interpretation of \ourlogic terms is defined inductively at the top of
figure~\ref{fig:semantics}. For all three term types, the
interpretation of a literal is the value denoted by the literal
itself; a variable is interpreted using the variable assignment
function for the corresponding type; an arithmetic expression defined
using a function $f$ is
interpreted by applying the interpretation of the function symbol $f$
to the interpretation of the corresponding arguments. The
operators \lit{i2t}, \lit{t2i}, \atst,  and \attt\ are
interpreted according to the informal semantics provided in the previous
section. 

The semantics of an \ourlogic formula $\phi$ is defined over a trace $\pi$
and a variable assignment $\mu$; we use the notation
$(\pi, \mu) \models \phi$ to indicate that trace $\pi$ satisfies formula
$\phi$ under variable assignment $\mu$. The satisfiability relation of
\ourlogic formulae is defined inductively at the bottom of
figure~\ref{fig:semantics}.  The formula
$ \texttt{tm}_1 < \texttt{tm}_2$ is satisfied if and only if (iff) the
interpretation of term $\texttt{tm}_1$ is lower than the
interpretation of term $\texttt{tm}_2$.  The semantics of the Boolean
connectives \lit{or} and \lit{not} is the standard one. A formula with
an existential quantifier over a timestamp variable, of the form
$\lit{exists}\ \tau\ \lit{in}\ I_T\ \lit{such that}\ \synt{p}$, is
satisfied iff there exists a timestamp $t_j \in I_T$, such that when
substituting timestamp $t_j$ for $\tau$ in the formula \synt{p}
(denoted by $ \synt{p}[\tau \leftarrow t_j]$), the resulting formula is
satisfied.  The semantics of the other two types of formulae with an existential quantifier is defined in a similar way.

\begin{figure}
\centering
\footnotesize
\begin{tabular}{p{0.88\columnwidth}}
\toprule
\emph{Time Term Interpretation}\\ \midrule
$\interpretation{\tau}_{\pi, \mu} = \mu^\mathit{TV}(\tau)$, for all
$\tau \in \mathit{TV}$;\\ $\interpretation{t}_{\pi, \mu}  =  t$, for all $t \in \mathbb{T}$;\\
$\interpretation{\lit{i2t}(\synt{it})}_{\pi, \mu}  = \pi[\interpretation{\synt{it}}_{\pi, \mu}].t $;\\
$\interpretation{f(\synt{tt}_1, \synt{tt}_2)}_{\pi, \mu}  =  \interpretation{f}_{\pi, \mu}(\interpretation{\synt{tt}_1}_{\pi, \mu} ,\interpretation{\synt{tt}_2}_{\pi, \mu})$;\\
\midrule \emph{Index Term Interpretation}\\ \midrule
$\interpretation{\sigma}_{\pi, \mu} = \mu^\mathit{IV}(\sigma)$, for all
$\sigma \in \mathit{IV}$;\\ $\interpretation{j}_{\pi, \mu}  =  j$, for all $j \in \mathbb{J}$;\\
$\interpretation{\lit{t2i}(\synt{tt})}_{\pi, \mu}  = \iota_\pi(\interpretation{\synt{tt}}_{\pi, \mu})$;\\
$\interpretation{f(\synt{it}_1, \synt{it}_2)}_{\pi, \mu}  =  \interpretation{f}_{\pi, \mu}(\interpretation{\synt{it}_1}_{\pi, \mu} ,\interpretation{\synt{it}_2}_{\pi, \mu})$;\\
\midrule\emph{Value Term Interpretation}\\  \midrule
$\interpretation{\rho}_{\pi, \mu} = \mu^\mathit{RV}(\rho)$, for all
$\rho \in \mathit{RV}$;\\ $\interpretation{x}_{\pi, \mu}  =  x$, for all $x \in \real$;\\
$\interpretation{(s \atst \synt{it})}_{\pi, \mu}  =
\pi[\interpretation{\synt{it}}_{\pi, \mu}].s$;\\ 
$\interpretation{(s \attt \synt{tt})}_{\pi, \mu}  = \pi[\iota_{\pi}(\interpretation{\synt{tt}}_{\pi, \mu})].s$\\
$\interpretation{f(\synt{vt}_1, \synt{vt}_2)}_{\pi, \mu}  =  \interpretation{f}_{\pi, \mu}(\interpretation{\synt{vt}_1}_{\pi, \mu} ,\interpretation{\synt{vt}_2}_{\pi, \mu})$;\\
\end{tabular}
\begin{tabular}{lll}
\midrule
\multicolumn{3}{c}{\textit{Formula Satisfaction}}\\ \midrule
$(\pi, \mu) \models \synt{tm}_1 < \synt{tm}_2$   & iff &
$\interpretation{\synt{tm}_1}_{\pi, \mu} < \interpretation{\synt{tm}_2}_{\pi, \mu}$\\ 
$(\pi, \mu) \models \lit{not}\ \synt{p}$ & iff & $(\pi, \mu) \not \models \synt{p}$\\
$(\pi, \mu) \models \synt{p}_1\ \lit{or}\  \synt{p}_2$ & iff & $(\pi, \mu) 
\models \synt{p}_1 \text{ or } (\pi, \mu) \models \synt{p}_2$\\
$(\pi, \mu) \models \lit{exists}\ \tau\ \lit{in}\ I_T$ & iff & $(\pi, \mu) \models \synt{p}[\tau \leftarrow t_j]$ \\
\phantom{$(\pi, \mu) \models$}$\;\lit{such that}\  \synt{p}$ &  & \phantom{$(\pi, \mu) \models$}  for some $t_j \in I_T$\\
$(\pi, \mu) \models \lit{exists}\ \sigma\ \lit{in}\ I_J$ & iff & $(\pi, \mu) \models \synt{p}[\sigma \leftarrow j]$ \\
\phantom{$(\pi, \mu) \models$}$\;\lit{such that}\  \synt{p}$ &  & \phantom{$(\pi, \mu) \models$}  for some $j \in I_J$\\
$(\pi, \mu) \models \lit{exists}\ \rho$ & iff & $(\pi, \mu) \models \synt{p}[\rho \leftarrow v]$ \\
\phantom{$(\pi, \mu) \models$}$\;\lit{such that}\  \synt{p}$ &  & \phantom{$(\pi, \mu) \models$}  for some $v \in \real$\\
\bottomrule  
\end{tabular}

 \caption{Semantics of the Hybrid Logic of Signals.}
\label{fig:semantics}
\end{figure}

\section{Trace Checking HLS formulae}
\label{sec:Theodore}
In this section, we present \app, our trace checker for \ourlogic.
\app reduces the problem of checking an \ourlogic property on a trace to a
satisfiability problem, which can be solved  using off-the-shelf
SMT solvers.

\app takes as input a property $\phi$ expressed in \ourlogic and a
trace $\pi$. The first step of \app is to automatically translating property $\phi$  and trace
  $\pi$  formulae expressed using a target
  logic $\mathcal{L}$. This translation relies on two  translation
  functions $\translation$ (for \ourlogic formulae, see Section~\ref{sec:formula2SMT}) and
  $\tracetrans$ (for traces, see Section~\ref{sec:translationtrace}) and guarantees, that $(\pi,\mu) \models \phi \text{ \textbf{iff} } \translation(\neg \phi)\wedge \tracetrans(\pi) \text{ is not satisfiable}  $,  where $\mu$ is a model for $\translation(\neg \phi)\wedge \tracetrans(\pi)$), i.e., $\mu$ is a variable assignment leading to the property violation, consistent with the values of the variables of the trace records. 

The second step of \app is checking the satisfiability of formula  $\psi \equiv \translation(\neg
  \phi) \wedge \tracetrans(\pi)$, expressed in the target
  logic $\mathcal{L}$ using an SMT solver.
Based on the condition stated above, when $\psi$ is satisfiable, it means
that $\phi$ does not hold on the trace $\pi$. Vice-versa, when $\psi$
is not satisfiable, it means that $\phi$ holds on the trace $\pi$.

The final verdict yielded by \app  can be ``\emph{satisfied}'', ``\emph{violated}'' or ``\emph{unknown}''; it is based on the answer of the solver. 
\app yields the \emph{definitive verdicts} ``satisfied'' or ``violated''  when the solver returns ``UNSAT'' or ``SAT'', indicating, respectively, that $\psi$ is unsatisfiable or satisfiable.
However, the solver may return an ``UNKNOWN'' answer, since the
satisfiability of the underlying target logic $\mathcal{L}$ is
generally undecidable. In our case, this indicates that no conclusion
is drawn on the satisfiability of formula $\psi$, resulting in an
``unknown'' verdict returned by \app. Assessing whether this is a frequent
case in practical applications is part of our evaluation
(Section~\ref{sec:evaluation}).

The target logic $\mathcal{L}$ to be selected for trace checking of
\ourlogic properties in \app shall fulfill two goals:
\begin{asparaenum}[\itshape {G}1:]
\item be sufficiently expressive to encode the logic-based
  representation of an input trace $\pi$ and the (semantics of an)
  \ourlogic formula $\phi$. This means that it should include linear
  real arithmetic (to support real-valued and timestamp terms),
  quantifiers (since \ourlogic is a first-order logic), and arrays
  (since a trace can be seen as an array of records).
  
\item be supported by an efficient solver, so that the
  trace checking procedure for \ourlogic formulae can be completed
  within practical time limits.
\end{asparaenum}

We have identified the \emph{AUFLIRA} (Closed linear formulae with
free sort and function symbols over one- and two-dimentional arrays of
integer indices and real values) fragment of the SMT-LIB
(Satisfiability Modulo Theories LIBrary) logic~\cite{barrett:the-smt-lib-sta} as a
suitable target logic for \app. The theories used by AUFLIRA are
identifiable through its name: \underline{A}: arrays; \underline{UF}:
extension allowing free sort and function symbols; \underline{LIRA}:
linear integer and real arithmetics. Furthermore, AUFLIRA does not
restrict the formulae to be quantifier-free. Based on the list of
supported theories, AUFLIRA satisfies G1. It also satisfies G2, since
it is included in the SMT-LIB logic, whose  satisfiability can be
verified using  highly efficient and optimized solvers, as shown in the annual SMT competition~\cite{WCDHNR19}.

In the following subsections we will describe functions 
\tracetrans\ and \translation. 
For simplicity, we will present the translation using the syntax of the 
Z3 Python API~\cite{z3online}.

\subsection{Translating a Trace into the Target Logic}
\label{sec:translationtrace}
Function \tracetrans\ translates a trace $\pi$ into a logic formula
expressed using the target logic $\mathcal{L}$.

To represent the sequence of timestamps in $\pi$, the translation
 creates an array variable \lstinline!t!; the type of the array
indices (i.e., the domain of \lstinline!t!) is $\mathbb{Z}$, whereas
the type of the array values (i.e., the range of \lstinline!t!) is $\mathbb{R}$. Then, the translation defines a series of constraints on
the values in \lstinline!t!: the value of array
\lstinline!t! at position $i$ (denoted by
\lstinline[mathescape]!t[$i$]!) is constrained to be equal to the
value of the timestamp contained in the record at index $i$ of trace
$\pi$.

In addition, the translation creates an array variable \emph{for each
  signal} whose values are recorded in the trace; the variable name is
the string obtained by concatenating \texttt{v\textunderscore} with the 
name of the signal. For each of these array variables
representing signals, the translation defines a series of constraints
on the values of the array: the value of the array in position $i$ is
constrained to be equal to the value of the corresponding signal in
the record at index $i$ of trace $\pi$.

 \subsection{Translating an HLS Formula into the Target Logic}
\label{sec:formula2SMT}

Function \translation\ translates an \ourlogic formula into a  logic
formula expressed using the target logic $\mathcal{L}$.

First, the translation declares a new variable for each
timestamp, index, and real-valued variable used in the HLS formula;
the name of the new variable is the string obtained by concatenating
\texttt{v\textunderscore} with the named of the original variable.
The type of the new variables is \lstinline!Real! for timestamp and
real-valued variables, and \lstinline!Int! for index variables.

Afterwards, the translation recursively evaluates each node in the
parse tree of the input formula, starting from the root node; each
node is translated using the rules shown in
\figurename~\ref{tab:HLS2TL}.

The translation of time, index, and values term nodes is defined as
follows. Nodes referring to \ourlogic variables are translated into
the corresponding variables in the target logic formula. Literal nodes
are mapped into literals in the target logic
formula. Arithmetic expressions using a function $f$ are translated by
converting the function symbol into the equivalent in the target
language, and then by applying it to the translation of its arguments.
A time term node of the form $\lit{i2t}(\synt{it})$ is translated into
an expression that accesses the element of the array \lstinline!t! in
position $\translation(\synt{it})$. An index term node of the form
$\lit{t2i}(\synt{tt})$ is translated into the application of the
translation of function
$\iota$ to $\translation(\synt{tt})$. A value term of the form
$(s \atst \synt{it})$ is translated into an expression that retrieves
the value of variable \texttt{v\textunderscore$s$} at index
$\translation(\synt{it})$. Similarly, a value term of the form
$(s \attt \synt{tt})$ is translated into an expression that retrieves
the value of variable \texttt{v\textunderscore$s$} at the index
obtained through the evaluation of
$\translation(\iota)(\translation(\synt{tt}))$.

The translation of function $\iota$ supports both definitions
presented in section~\ref{sec:bkg:traces}. It consists of a rewriting
of the definition into the equivalent syntax of the target logic.  We
remark that the size of the arithmetic expression to compute
$\translation(\iota^V)$ in the case of a variable sample rate is
linear in the length of the trace and the number of timestamp
variables. Evaluating the impact of our translation and of the
selection of the definition of function $\iota$ on the performance of
the trace-checking procedure is part of our evaluation.

The translation of \ourlogic formulae is basically their rewriting
into the equivalent syntax of the target logic, modulo the translation
of the variables and of the sub-formulae. For example, a formula of
the form $\lit{exists}\ \sigma\ \lit{in}\ I_J\ \lit{such that}\ \synt{p}$ is rewritten
as 
$\texttt{Exists}\left(\texttt{v\textunderscore}\sigma,
  \texttt{And}\left(\texttt{And}\left(a \leq \texttt{v\textunderscore}\sigma,
  \texttt{v\textunderscore}\sigma \leq b\right), \translation\left(\synt{p}\right)\right)\right)$,
where the target logic variable $\texttt{v\textunderscore}\sigma$
corresponds to variable $\sigma$ in the \ourlogic formula,
$a$ and $b$ are the lower and upper bounds of the closed interval $I_j$, and
$\translation(\synt{p})$ is the translation of sub-formula \synt{p}.\footnote{Our translation also supports open intervals. In this case, the relational operator $<$ (instead of $\leq$) is used in the target logic formula to constrain the values
$\texttt{v\textunderscore}\sigma$ can assume.}

\begin{figure}[t]
\centering
\footnotesize
\begin{tabular}{l}
  \toprule
\emph{Time Term}\\ \midrule
$\translation(\tau)= \texttt{v\textunderscore}\tau$, for all $\tau \in \mathit{TV}$;\\ 
$\translation(t)=t$, for all $t \in \mathbb{T}$;\\
$\translation(f(\synt{tt}_1, \synt{tt}_2)) =
  \translation(f)(\translation(\synt{tt}_1), \translation(\synt{tt}_2))$;\\ 
$\translation(\lit{i2t}(\synt{it})) =
  \texttt{t[}\translation(\synt{it})\texttt{]}$; \\
\midrule \emph{Index Term}\\ \midrule
$\translation(\sigma)= \texttt{v\textunderscore}\sigma$, for all $\sigma \in \mathit{IV}$;\\
$\translation(j)=j$, for all $j \in \mathbb{J}$;\\
$\translation(f(\synt{it}_1, \synt{it}_2)) =
  \translation(f)(\translation(\synt{it}_1), \translation(\synt{it}_2))$;\\
$\translation(\lit{t2i}(\synt{tt})) =
  \translation(\iota)(\translation(\synt{tt}))$; \\
\midrule\emph{Value Term}\\  \midrule
$\translation(\sigma)= \texttt{v\textunderscore}\sigma$, for all $\sigma \in
  \mathit{RV}$;\\ $\translation(x)=x$, for all $x \in \real$;\\
$\translation(f(\synt{vt}_1, \synt{vt}_2)) =
  \translation(f)(\translation(\synt{vt}_1), \translation(\synt{vt}_2))$\\       
$\translation((s \atst \synt{it})) = \texttt{v\textunderscore
 s[}\translation(\synt{it})\texttt{]}$;\\
$\translation((s \attt \synt{tt})) = \texttt{v\textunderscore
 s[}\translation(\iota)\left(\translation\left(\synt{tt}\right)\right)\!\texttt{]}$;\\
\midrule\emph{Formula} (with $I_T=[t_a,t_b]$ and $I_J=[a,b]$)\\  \midrule
$\translation(\synt{tm}_1 < \synt{tm}_2) = \translation(\synt{tm}_1) <
  \translation(\synt{tm}_2)$; \\
$\translation(\synt{p}_1\ \lit{or}\  \synt{p}_2) =
  \texttt{Or}(\translation(\synt{p}_1), \translation(\synt{p}_2))$;\\ 
  $\translation(\lit{not}\ \synt{p}) = \texttt{Not}(\translation(\synt{p}))$; \\
$\translation( \lit{exists}\ \tau\ \lit{in}\ I_T\ \lit{such that}\ 
  \synt{p}) = $\\
  $\quad \texttt{Exists}\left(\texttt{v\textunderscore}\tau,
  \texttt{And}\left(\texttt{And}\left(t_a \leq \texttt{v\textunderscore}\tau,
  \texttt{v\textunderscore}\tau \leq t_b\right), \translation\left(\synt{p}\right)\right)\right)$  \\
$\translation(\lit{exists}\ \sigma\ \lit{in}\ I_J\ \lit{such that}\
  \synt{p}) = $\\
  $\quad \texttt{Exists}\left(\texttt{v\textunderscore}\sigma,
  \texttt{And}\left(\texttt{And}\left(a \leq \texttt{v\textunderscore}\sigma,
  \texttt{v\textunderscore}\sigma \leq b\right), \translation\left(\synt{p}\right)\right)\right)$  \\
$\translation(\lit{exists}\ \rho\ \lit{such that}\  \synt{p}) =
  \texttt{Exists}(\texttt{v\textunderscore}\rho, \translation(\synt{p})) $\\
\bottomrule
\end{tabular}

 \caption{Rules for translating \ourlogic formulae into $\mathcal{L}$.}
\label{tab:HLS2TL}
\end{figure}

  \app ensures that $(\pi,\mu) \models \phi$ \textbf{iff}  $\translation(\neg \phi)\wedge \tracetrans(\pi)$  is not satisfiable. 
 The correctness of our procedure is based on two arguments:
\begin{inparaenum}[(i)]
\item \tracetrans\  translates the trace $\pi$ into a set of array variables whose values are set according to the values of the original trace, and
\item \translation\ rewrites the \ourlogic formula into the target logic without applying any change (that could alter the semantics) to the structure of the formula.
\end{inparaenum}

\subsection{Implementation}
\label{sec:implementation}
We implemented \app as an Eclipse plugin using Xtext~\cite{Xtext} and Xtend~\cite{xtend} and made it publicly available~\cite{ThEodorE,menghi_claudio_2021_4506796}. We selected Z3~\cite{z3online} as SMT solver, since it is an award-winning~\cite{sigplanAward, etapsAward}, industry-strength tool.
As such, it is likely to satisfy
 goal \emph{G2} discussed above. 
Checking whether this conjecture holds is part of our evaluation.

\section{Evaluation}
\label{sec:evaluation}
In this section, we report on the evaluation of our contributions.  First, we evaluate the expressiveness of \ourlogic, and compare it with state-of-the-art specification languages. 
Second, we evaluate the applicability of the \app trace checker, and compare it to  state-of-the-art tools.  Specifically, we aim to answer the following research questions:
\begin{compactenum}[{RQ}1]
\item \emph{To which extent can \ourlogic express requirements from 
  industrial CPS applications and how does it compare with 
  state-of-the-art specification languages in terms of
  expressiveness?} (section~\ref{sec:eval:rq1})
\item \emph{Can \app verify CPS requirements on real-world
  execution traces within practical time and how does it compare
  with  state-of-the-art tools?} (section~\ref{sec:eval:rq2})
  \end{compactenum}

  \subsection{Expressiveness of \ourlogic (RQ1)}
\label{sec:eval:rq1}
To answer RQ1, we  collected a set of  industrial CPS requirements expressed in plain English text, and verified whether they could be expressed in \ourlogic and in other state-of-the-art specification languages. 

\emph{Dataset}. We considered  \rqonerequirements\ industrial requirements from our satellite case study,  coming from three different sources:
\begin{asparaenum}[{$\mathcal{S}$}1:]
    \item $61$ requirements were randomly selected from $745$ requirements contained in the  requirement specification document of the satellite on-board software (OBSW). Due to the prohibitive effort (more than $20$ hours spanned across several working days) involved, both on our part and that of the domain experts who helped us formalize these requirements, we could only process a subset. 
Such requirements mostly refer to the software dynamics of the satellite, as in ``\emph{When the satellite switches to ``Idle Mode'', the OBSW shall checkout the GPS, wait \SI{50}{\milli\s}, and then checkout the sun sensors}''.
\item  $101$  requirements were provided by the authors of SB-TemPsy-DSL~\cite{ase2020}. They mostly refer to the physical dynamics of the satellite, as in ``\emph{the beta angle~\cite{beta} shall show an oscillatory behavior with a maximum period of \SI{2500}{\s}}''.
\item $50$  requirements were extracted from the design and architectural  documents of the satellite. 
These documents describe the relations and interactions among the different components of the satellite. 
They contain cyber-physical  requirements that relate the software and the physical dynamics of the satellite, as in ``\emph{if the satellite mode switches from ``Idle Mode'' to ``Safe Spin Mode'' and the satellite is not in eclipse, the magnetic field recorded by the magnetometer shall contain a spike with a maximum amplitude of \SI{0.02}{\tesla}}''.
\end{asparaenum}

\emph{Methodology.} We tried to express the requirements from our dataset using \ourlogic and two 
state-of-the-art specification languages, namely SB-TemPsy-DSL~\cite{ase2020} and 
STL~\cite{maler2004monitoring}. We selected these languages because they are both 
supported by trace checking tools. We assessed the extent to which requirements were expressible in each language.

\emph{Results.} Table~\ref{tab:exp} reports\footnote{The values in Table~\ref{tab:exp} marked with an asterisk are slightly different from those reported in~\cite{ase2020}. In the latter, 
quantification on  real-valued variables (not supported in STL and SB-TemPsy-DSL) was handled by artificially 
selecting a value  for the quantified variables within their quantification range. In this work, 
we marked such requirements as not specifiable.} the number of requirements that we were able to 
express in each of the languages, for each set of requirements ($\mathcal{S}1$, $\mathcal{S}2$, and $\mathcal{S}3$).
\ourlogic was able to express \rqoneHLSpercentage\% (\rqonerequirements/\rqonerequirements) of the requirements, while SB-TemPsy-DSL and STL were able to express \rqoneTempsypercentage\% (\rqoneTempsynumber/\rqonerequirements) and \rqoneSTLpercentage\% (\rqoneSTLnumber/\rqonerequirements) of the requirements, respectively. 
These results confirm that \ourlogic is highly expressive and much more so than alternatives. 
We remark that all the \ourlogic constructs  were useful to express at least some of the considered CPS requirements, though in very different proportions.

\resq{The answer to RQ1 is that \ourlogic could express \emph{all} the requirements of our case study,  many more than SB-TemPsy-DSL (\num{+\xinttheexpr \rqonerequirements -\rqoneTempsynumber\relax}) and STL (\num{+\xinttheexpr \rqonerequirements - \rqoneSTLnumber\relax}).  
}

\begin{table}[]
\caption{Number of requirements expressible in each of the languages for each set of requirements.}
    \label{tab:exp}
    \centering
    \begin{tabular}{l l l ll}
    \toprule
       & $\mathcal{S}1$ & 
       $\mathcal{S}2$ & $\mathcal{S}3$ & Total \\
       \midrule
        \ourlogic  & $61$/$61$ & $101$/$101$ & $50$/$50$ & $212/212\ (100\%)$ \\
        SB-TemPsy-DSL & $34/61$ &  $92$/$101$$^\ast$ & $19/50$ & $145/212\ (68\%)$\\
        STL & $38/61$ & $51$/$101$$^\ast$ & $13/50$  & $102/212\ (48\%)$ \\
    \bottomrule
    \end{tabular}\\

\end{table}

\subsection{Applicability of \app (RQ2)}
\label{sec:eval:rq2}
To answer RQ2, we
\begin{inparaenum}[(i)]
\item assessed to which extent \app can be applied to check the execution traces of our case study;
\item compared, in terms of applicability, \app with SB-TemPsy-Check~\cite{ase2020} and Breach~\cite{DBLP:conf/cav/DonzeFM13}.
\end{inparaenum}
SB-TemPsy-Check is the  trace checker for SB-TemPsy-DSL; 
Breach is a trace checker for STL. 
We chose Breach among other similar tools listed in a recent survey~\cite{bartocci2018specification} (i.e., AMT~\cite{nickovic2018:amt-2.0:-qualit,nivckovic2020amt} and S-TaLiRo~\cite{annpureddy2011s}), because AMT 2.0, in contrast to Breach, is not publicly available, and  because Breach is faster than S-TaLiRo~\cite{DBLP:conf/cav/DonzeFM13}. Furthermore, we excluded from our comparison tools for online trace checking (e.g., SOCRaTEs~\cite{FSE2019} and RTAMT~\cite{nickovic20:_rtamt}).

\textit{Dataset.} Our industrial partner provided  \numsimulationtraces\ traces,  obtained by simulating  the behavior of the satellite in different scenarios; the simulation time ranged from four to six hours. 
 Their
size (in number of entries) ranges from \minnumberofentries\ to
\maxnumberofentries\ entries ($\mathit{avg}=\averageumberofentries$, $
\mathit{sd}=\stdevnumberofentries$); the corresponding file size ranges from
\SI{\approx\minnumberofmem}{\mega\byte} to
\SI{\approx\maxnumberofmem}{\mega\byte}
($\mathit{avg}\ \SI{\approx\averagenumberofmem}{\mega\byte}$, $
\mathit{sd}\ \SI{\approx\stdevnumberofmem}{\mega\byte}$).
The traces have a considerably large (yet variable) number of records and size.

For each trace in our dataset, our industrial partner indicated which requirements to check.
Indeed, since only a subset of the satellite signals is recorded in each simulation scenario,
not all the requirements have to be checked on each trace.
In total, we considered \tracerequirementcombinations\ trace-requirement combinations: 
\tracerequirementcombinationsone\ obtained from requirements in $\mathcal{S}1$,
\tracerequirementcombinationstwo\ obtained from requirements in $\mathcal{S}2$,
and \tracerequirementcombinationsthree\ obtained from traces in $\mathcal{S}3$.
We remark that, out of these \tracerequirementcombinations\ combinations, \numexperimentsonlyHLS\ involve a requirement that can be expressed neither in SB-TemPsy-DSL  nor in STL.

Our industrial partner used a variable sample-rate for generating the trace records; hence  not  all the signal values were recorded at each sample index. 
Since our approach assumes that all the signals are assigned a value at each sample index, we pre-processed the traces.
First,  for each trace-requirement combination, we filtered out from the trace all the records that contained only
signals that were not used in the \ourlogic  specification of the requirement.
This step prevents the trace checker from handling an unnecessarily large set of records.
Then, we transformed the  traces using both pre-processing strategies $\mathcal{A}1$  and $\mathcal{A}2$ presented in section~\ref{sec:bkg:traces}; in both cases, the interpolation function to use for each signal was indicated by the engineers of our industrial partner.

By applying the $\mathcal{A}1$ and $\mathcal{A}2$ strategies on the original \tracerequirementcombinations\ trace-requirement combinations, the 
  final dataset contains  \tracerequirementcombinationstwoencodings\  trace-requirement combinations (with half of them obtained using one of the two strategies).
The size of the traces obtained using $\mathcal{A}1$  ranges from~\aoneminnumberofentries\ to
\aonevmaxnumberofentries\ entries ($\mathit{avg}=\aoneaverageumberofentries$, $\mathit{sd}=\aonestdevnumberofentries$); 
the corresponding file size ranges from \SI{\approx\aoneminmem}{\byte} to
\SI{\approx\aonemaxmem}{\mega\byte}
($\mathit{avg}\ \SI{\approx\aoneavgmem}{\mega\byte}$, $\mathit{sd}\ \SI{\approx\aonestdmem}{\mega\byte}$).
The size of the traces obtained using $\mathcal{A}2$ ranges from \atwominnumberofentries\ to
\atwomaxnumberofentries\ entries ($\mathit{avg}=\atwoaverageumberofentries$, $\mathit{sd}=\atwostdevnumberofentries$); the  file size ranges  from
\SI{\approx\atwominmem}{\byte} to
\SI{\approx\atwomaxmem}{\mega\byte}
($\mathit{avg}\ \SI{\approx\atwoavgmem}{\mega\byte}$, $\mathit{sd}\ \SI{\approx\atwostdmem}{\mega\byte}$).

\emph{Methodology.} We ran \app over the \tracerequirementcombinationstwoencodings\ trace-requirements combinations 
in our dataset. When translating the \ourlogic properties in the target logic, we used 
function $\iota^{V}$ for the trace-requirement combinations generated using strategy $\mathcal{A}1$ (since the pre-processed traces have a variable 
sample rate), and function $\iota^{F}$ for those generated using strategy $\mathcal{A}2$ (since the pre-processed traces have a fixed sample rate).

We conducted our evaluation on a high-performance computing platform, using nodes equipped with
Dell C6320 units (2 Xeon E5-2680v4@\SI{2.4}{\giga\hertz}, \SI{128}{\giga\byte}).\footnote{
We executed our experiments on the HPC facilities of the University of Luxembourg~\cite{VBCG_HPCS14}.} 
Each run (checking a distinct combination of a trace and a property) was repeated 10 times, to account for variations in the performance of the HPC platform and of the SMT solver. 
In total, we executed  $\tracerequirementcombinationstwoencodings \times 10 = \totalexperiments$ runs of \app. 
We allocated $\SI{4}{\giga\byte}$ of memory for each run and  considered a timeout of one hour. We recorded whether the trace-checking procedure ended 
within the timeout, the trace checking result, and the time required to yield a verdict.

As for the comparison with SB-TemPsy-Check and Breach, we only considered the requirements from $\mathcal{S}2$ since 
it has the highest number of requirements expressible in SB-TemPsy-DSL and STL, and it was  recently used for comparing SB-TemPsy-DSL with STL~\cite{ase2020}. More specifically, we considered 
the \thedoretempsycomparisontraced\ trace-requirement combinations (with requirements from the  set~$\mathcal{S}2$) expressible in SB-TemPsy-DSL, and the \thedorebreachcomparisontraced\ trace-requirement combinations expressible in STL.
We ran the tools  following the same methodology described above.
Since each run was repeated ten times, in total we considered 
\thedoretempsycomparisontracedexperiments\ runs of SB-TemPsy-Check and 
\thedorebreachcomparisontracedexperiments\  runs of Breach.

 \begin{table}[t]
 \caption{
Output of \app (percentage and execution time) when using the pre-processing strategies $\mathcal{A}1$ and $\mathcal{A}2$.}
     \label{tab:rq2a}
     \centering
     \begin{tabular}{l l l  l l  l  l  }
     \toprule
& \textbf{Output} & $\mathbf{\%}$ & 
    $\mathit{avg}$
      & $\mathit{min}$ & $\mathit{max}$ & $\mathit{sd}$ \\
      \midrule
 \multirow{6}{*}{$\mathcal{A}1$} &       \emph{satisfied}   & 
      \aonepositiveperc
        & \aonepositiveavg 
         & \aonepositivemin
        &   \aonepositivemax
        & \aonepositivedev
       \\
       & \emph{violated}  &  
        \aonenegativeperc &
\aonenegativeavg &
\aonenegativemin &
\aonenegativemax &
\aonenegativedev 
        \\  
       & \emph{unknown} 
        & 
\aoneundecidedperc
& \aoneundecidedavg
& \aoneundecidedmin
& \aoneundecidedmax
 &\aoneundecideddev 
        \\  
       & \emph{timeout} & \aonetimeout & - & - & - & - \\
      &  \emph{max\_depth\_exceeded} & \aonemaxdepth & - & - & - \\
       & \emph{out\_of\_memory} & \aonoutofMemory & - & - & - \\
    \midrule
\multirow{5}{*}{$\mathcal{A}2$}   & \emph{satisfied}   & 
      \atwopositiveperc
        & \atwopositiveavg 
         & \atwopositivemin
        &   \atwopositivemax
        & \atwopositivedev
       \\
&       \emph{violated}  &  
        \atwonegativeperc &
\atwonegativeavg &
\atwonegativemin &
\atwonegativemax &
\atwonegativedev 
        \\  
  &      \emph{unknown} 
        & 
\atwoundecidedperc
& \atwoundecidedavg
& \atwoundecidedmin
& \atwoundecidedmax
 &\atwoundecideddev 
        \\  
   &      \emph{timeout} & \atwotimeout & - & - & -& - \\
    \bottomrule
     \end{tabular}
      \end{table}

\emph{Results -  Applicability of \app.}
Table~\ref{tab:rq2a} shows the different types of output returned by \app for checking the \numexperimentsatwo\ trace-requirement combinations generated 
using the variable sample rate interpolation (row $\mathcal{A}1$) and the fixed sample rate interpolation (row $\mathcal{A}2$). Column ``\%'' indicates the percentage of cases in which each type of verdict was returned.
For each of the cases in which \app finished within the timeout (i.e., it yielded a \emph{satisfied}, \emph{violated}, or  \emph{unknown} verdict), Table~\ref{tab:rq2a} also provides the average ($\mathit{avg}$), minimum ($\mathit{min}$), maximum ($\mathit{max}$) and standard deviation ($\mathit{sd}$) of the \app execution time (\si{\s}).

The results in row $\mathcal{A}1$ show that \app finished within the timeout in \aonewithintimeoutpercentage\  of the cases.
In \aoneverdictintime\ of the cases, \app produced a definitive verdict (i.e., \emph{satisfied} or \emph{violated}); in \aonetimeout\% of the cases, \app 
timed out.  \app  returned a ``\emph{max\_depth\_exceeded - maximum recursion 
depth exceeded during compilation}'' error in $\aonemaxdepth\%$ of the cases, and an ``\emph{out\_of\_memory}'' error in $\aonoutofMemory\%$ 
of the cases; both errors are generated by the Z3 solver. The root cause of these errors is the translation of function $\iota^V$, used in the case of 
variable sample rate traces: the size of the arithmetic expression resulting from the translation is linear in the length of the trace.
 As expected, \app inherits the limitations of SMT solvers and its applicability is expected to improve along with the quick pace of progress in that field. 
 
The results in row $\mathcal{A}2$ show that \app finished within the timeout in \atwowithintimeoutpercentage\% of the cases. 
In $74.5$\% of the cases, \app produced a definitive verdict; in \atwotimeout\% of the cases, \app timed out. When using strategy $\mathcal{A}2$,
the number of times \app reached the timeout was higher than when using $\mathcal{A}1$. Indeed, many trace-requirement runs that generated \emph{max\_depth\_exceeded} and \emph{out\_of\_memory} errors in the case of 
$\mathcal{A}1$, timed out when using $\mathcal{A}2$.  As discussed for the case of $\mathcal{A}1$, the applicability of \app when using $\mathcal{A}2$ is determined by the scalability of the underlying SMT solver. 

To evaluate whether \app is applicable in cases in which neither SB-TemPsy-Check nor Breach is applicable, 
we considered the subset of \numrunsonlyHLS\ runs associated with the \numexperimentsonlyHLS\  trace-requirement combinations   that involve a requirement that can be expressed neither in SB-TemPsy-DSL  nor in STL.
For those combinations, \app was able to produce a verdict in \onlyHLSverdicpercentage\ of the cases.

To evaluate the impact of the trace accuracy (as determined by the application of the pre-processing strategies $\mathcal{A}1$ and $\mathcal{A}2$) on the correctness of the trace-checking procedure, 
we considered the $449$  runs in which \app returned a definitive verdict both when using $\mathcal{A}1$ and when using $\mathcal{A}2$, and we compared the verdicts.
In 95.1\% of the cases ($427$ over $449$), the verdicts coincided. 
For the $22$ cases in which the verdicts were different, we manually inspected the generated traces and confirmed that differences in verdicts were caused by the pre-processing strategies.
 
 Overall, these results show that \app, when configured with the pre-processing strategy based on a fixed sample rate ($\mathcal{A}2$), 
 produced a definitive verdict for a considerable number of trace-requirement combinations ($74.5$\%), thus confirming \app's applicability in practical scenarios.
Relying on the $\mathcal{A}2$ strategy led to a significantly wider applicability  of \app than with the $\mathcal{A}1$ strategy ($74.5$\%~vs~$66.0$\%), while resulting in negligible differences in trace accuracy.
Therefore, for comparing \app with other tools, we resorted to using the 
$\mathcal{A}2$ pre-processing strategy.

Finally, we remark that \app detected an issue in the satellite design: some of the traces exhibited an unexpected spike in a signal related to the physical dynamics of the satellite, which was caused by a change in a signal related to its software dynamics.

 \emph{Results - Comparison with other tools.} 
 Table~\ref{tab:comparison} reports the percentage of cases in which \app, SB-TemPsy-Check, and Breach provided a verdict within the timeout and  the minimum, maximum, average  and standard deviation of the  time required to yield the verdict.
 
 The results show that, when the requirements are expressible in SB-TemPsy-DSL and STL, SB-TemPsy-Check and
Breach are faster than \app. However, given the usage scenario considered in our work (offline trace checking), the difference in execution times reported in  Table~\ref{tab:comparison} does not have significant practical 
consequences since  the average trace-checking time (less than two minutes) is significantly lower than the time required to collect the traces (several hours). Note that all tools were consistent in terms of verdicts: when \app 
returned a definitive verdict, it matched the verdict returned by SB-TemPsy-Check and Breach (when they did not time out).

\resq{The answer to RQ2 is that \app could compute a definitive verdict, within one hour, for \percentagedefinitiveatwo\% of the trace-requirement combinations of our industrial case study, and 
 produced a verdict for \onlyHLSverdicpercentage\ of the  \numexperimentsonlyHLS\ trace-requirement combinations that could not be checked by the other tools.
 }

 \begin{table}[t]
 \caption{Comparison of \app, SB-TemPsy-Check, and Breach in terms of the execution time.}
     \label{tab:comparison}
     \centering
     \begin{tabular}{l l l  l  l  l}
     \toprule
     \textbf{Tool} &
$\mathbf{\%}$ & 
    $\mathit{avg}$
      & $\mathit{min}$ & $\mathit{max}$ & $\mathit{sd}$ \\
      \midrule
      \app   & 
       \Theodorevstempsypercentage &
\Theodorevstempsyavg &
\Theodorevstempsymin &
\Theodorevstempsymax &
\Theodorevstempsystd
       \\
        SB-TemPsy  &
        \sbtempsypercentage &
\sbtempsyavg &
\sbtempsymin &
\sbtempsymax &
\sbtempsystd
        \\ 
     \midrule
       \app   &  
       \Theodorevsbreachpercentage &
\Theodorevsbreachavg & 
\Theodorevsbreachmin &
\Theodorevsbreachmax &
\Theodorevsbreachstd
       \\
        Breach  &
         \breachpercentage &
\breachavg &
\breachmin &
\breachmax &
\breachstd
        \\  
        \bottomrule
     \end{tabular}
      \end{table}
 
 \subsection{Discussion and Threats to Validity}
 \label{sec:discussion}
 Based on results, we recommend the following workflow. 
 Developers should initially use \app since its language (\ourlogic) is the most expressive, and it is generally difficult to know in advance which requirement types engineers will need to specify. 
If the property to be verified does not contain the \lit{t2i} \ourlogic operator, which causes the generation of large arithmetic expressions, engineers  should use \app with the pre-processing strategy based on a variable sample rate ($\mathcal{A}1$). 
 If the property contains the  \lit{t2i} operator, 
 engineers should use the pre-processing strategy based on a fixed sample rate ($\mathcal{A}2$). If \app was not able to produce a definitive verdict, and the requirement is expressible in SB-TemPsy-DSL or STL, engineers should use SB-TemPsy-Check or Breach.
 
 \emph{Threats to validity.} 
 The requirements and traces we used in our evaluation come from a single case study in the
 satellite domain. Although 
 this could influence the generalization of our results, our  industrial case study is representative of what can be found in other cyber-physical domains, where the system requirements are complex properties related to the software system, its environment and their interactions, and traces are obtained by simulating (or executing) the behavior of the CPS in many different scenarios.

\section{Related Work}
\label{sec:related}
Our contribution is mainly related to work done in the area of hybrid specification  languages.

STL-MX~\cite{STLmx} extends STL to define properties both on discrete time  and on dense time. The language includes two layers, one based on LTL to express properties of discrete-time Boolean signals (sampled at a fixed sample rate), and another one based on STL, to express properties on 
dense-time real-valued signals. Time mapping operators define the conversion between dense-time and discrete-time signals and formulae. 
A trace-checking procedure has been proposed for STL-MX, but its implementation is not available. Compared with \ourlogic, STL-MX restricts 
discrete-time Boolean signals to be sampled at a fixed sample rate, and lacks first-order quantifiers on real-valued variables.

HyLTL~\cite{bresolin2013hyltl}, HRELTL~\cite{CIMATTI201554}, and  HTL~\cite{henzinger1992towards} 
extend existing languages (e.g., LTL) with operators to  express constraints on certain behaviors of signals (e.g., derivatives or limits).
In contrast to \ourlogic, they cannot express  properties that refer to specific time instants and to the distance between them.

Differential Dynamic Logic~\cite{platzer2008differential} differs from HLS since it is designed for specifying properties of systems expressed using the hybrid system~\cite{alur2015principles} modeling formalism. 
As such, its modal operators enable references to the states that are reachable after firing the transitions of the hybrid system model.

The approach of reducing the trace-checking problem to the verification of the satisfiability of a logical formula has been also used in other works (\cite{CIMATTI201554,10.1007/978-3-642-54804-8_19,bgks-soca2014}). However, our approach supports \ourlogic, a more widely applicable language, and developed an efficient translation for it.

SOCRaTEs~\cite{FSE2019}, Striver~\cite{gorostiaga2018striver},
TeSSLa~\cite{convent2018tessla}, and
RTLola~\cite{10.1007/978-3-319-46982-9_10} and a tool recently proposed by \citet{arrieta2020tool} are also related to our work.
Unlike ThEodorE, which supports offline trace checking, these tools support online run-time verification.

To summarise, in our context and given our goal, in addition to the lack of trace-checking tools, none of the languages discussed above is as expressive as \ourlogic. 
Taking into account the expressiveness limitations of state-of-the-art languages like SB-TemPsy-DSL and STL, which were not able to express many of our requirements (see section~\ref{sec:evaluation}), the development of a new language (and of the corresponding trace-checking tool) was indeed necessary.

\section{Conclusion}
\label{sec:conclusion}
Software verification and validation requires specification-driven trace-checking techniques that strike a balance between the expressiveness of the specification language and the efficiency of its trace-checking procedures. 
In this paper, we specifically address this problem in the CPS domain. We proposed the Hybrid Logic of Signals (\ourlogic), a specification language tailored to the specifics of CPS requirements. \ourlogic allows engineers to specify complex CPS requirements related to its cyber and the physical components, as well as their interactions.
Additionally, we developed \app, an efficient SMT-based trace-checking procedure for  \ourlogic.

We evaluated our solutions through a large-scale, complex industrial case study involving an on-board satellite system. 
Results show that our approach achieves a better trade-off between expressiveness and performance than existing solutions.
\ourlogic was able to express all system requirements in contrast to existing languages. 
As a result, \app supports a much wider set of property types than other trace checkers.
In most cases, \app  was able to check those properties within practical time limits. Furthermore, the applicability of \app is expected to improve in the future 
along with the underlying SMT technology.
Last, based on results, we suggest a way to effectively combine various trace-checking tools.

As part of future work, we plan to develop trace diagnostics methods for \ourlogic, inspired by existing work~\cite{dbb:models2018,ferrere2015:trace-diagnosti},
to explain the violations found by \app.
 
\section{Data Availability}
\app is publicly available~\cite{ThEodorE,menghi_claudio_2021_4506796} under the Apache License 2.0.
The entry on Zenodo.org~\cite{menghi_claudio_2021_4506796} contains, in addition to the software, the files containing the results produced by \app, and the scripts to compute the aggregated results presented in the paper.
The traces and requirements used in the experiments cannot be publicly released because they are subject to a non-disclosure agreement.

\section*{Acknowledgment}
  This work has received funding from the
  European Research Council
  under the European Union's Horizon 2020
  research and innovation programme (grant agreement
  No~694277),
  from the Natural Sciences and Engineering
    Research Council of Canada (NSERC)
  under the Discovery and CRC programs.
  
    The experiments presented in this paper were carried out
using the HPC facilities of the University of Luxembourg~\cite{VBCG_HPCS14}
{\small --- see \url{hpc.uni.lu}}.

\bibliographystyle{IEEEtranN}

\end{document}